\documentclass[runningheads,fleqn]{svmult}

\usepackage{makeidx} 
\usepackage{graphicx}  
\usepackage{subeqnar} 
\usepackage{multicol} 
\usepackage{physproc} 
\usepackage{amsmath}
\usepackage{amssymb}

\makeindex     

\begin{document}

\title*{New Insights from One-Dimensional Spin Glasses}
\toctitle{New Insights from One-Dimensional Spin Glasses}
\titlerunning{New Insights from One-Dimensional Spin Glasses}

\author{
     Helmut G.~Katzgraber\inst{1}
\and Alexander K.~Hartmann\inst{2}
\and A.~P.~Young\inst{3}
}
\authorrunning{Helmut G. Katzgraber et al.}
\institute{
     Theoretische Physik, ETH Z\"urich, CH-8093 Z\"urich, Switzerland
\and Institute for Physics, University of Oldenburg, Germany
\and Department of Physics, University of California Santa Cruz, CA 95064, USA
}

\maketitle 

\begin{abstract}
The concept of replica symmetry breaking found in the solution of
the mean-field Sherrington-Kirkpatrick spin-glass model has been
applied to a variety of problems in science ranging from biological
to computational and even financial analysis. Thus it is of paramount
importance to understand which predictions of the mean-field solution
apply to non-mean-field systems, such as realistic short-range
spin-glass models. The one-dimensional spin glass with random power-law
interactions promises to be an ideal test-bed to answer this question:
Not only can large system sizes---which are usually a shortcoming
in simulations of high-dimensional short-range system---be studied,
by tuning the power-law exponent of the interactions the universality
class of the model can be continuously tuned from the mean-field to
the short-range universality class. We present details of the model,
as well as recent applications to some questions of the physics of
spin glasses.  First, we study the existence of a spin-glass state
in an external field.  In addition, we discuss the existence of
ultrametricity in short-range spin glasses. Finally, because the range
of interactions can be changed, the model is a formidable test-bed
for optimization algorithms.
\end{abstract}

\section{Introduction}
\label{sec:introduction}

Spin glasses \cite{binder:86,mezard:87,young:98} are paradigmatic
models which can be applied to a wide variety of problems and fields
ranging from economical to biological, as well as sociological
problems, to name a few. Most prominent is the replica symmetry
breaking solution of Parisi \cite{parisi:79} of the mean-field
Sherrington-Kirkpatrick (SK) spin glass. Unfortunately, an analytical
solution for short-range realistic spin-glass models, such as the
Edwards-Anderson Ising spin glass \cite{edwards:75}, remain to
be found and generally phenomenological descriptions, such as the
droplet picture \cite{fisher:86} or numerical simulations are used
to understand these systems. Given the lack of rigorous results
for short-range spin glasses, it is of importance to understand the
applicability of different predictions made by the mean-field solution
of the SK model, as well as other theoretical pictures.

Unfortunately, numerical studies of spin glasses are difficult to
accomplish and in general only small to moderate system sizes can
be accessed.  Despite huge technological advances in the last decade
which have enabled the construction of large computer clusters out of
commodity components, brute force computation alone will not suffice
to probe considerably larger system sizes.  The source of this problem
lies in the diverging equilibration times of Monte Carlo simulations
of spin glasses; the systems are generally NP hard. Furthermore,
to obtain thermodynamically sound results, calculations need to
be disorder averaged, thus adding considerable overheard to any
simulation. To properly probe the thermodynamic limiting behavior
it is thus important to use efficient algorithms, improved models,
as well as large computer clusters.

We discuss in detail a one-dimensional spin-glass model with
power-law interactions \cite{fisher:86,kotliar:83,katzgraber:03}
where, by tuning the exponent of the power-law interactions different
universality classes from infinite-range SK to short-range can
be probed.  Furthermore, because the model is one-dimensional,
a wide range of system sizes can be probed. In the past we have
applied the model to different problems in the physics of spin glasses
\cite{katzgraber:03,katzgraber:03f,katzgraber:04c,katzgraber:05c,boettcher:07,katzgraber:08}.
In this work we study two questions which lie at the core of
the applicability of the mean-field solution to short-range spin
glasses: Do short-range spin glasses order in an externally-applied
magnetic field?  Are short-range spin glasses ultrametric? Our results
suggest that new theoretical descriptions are needed: While there are
indications of an ultrametric structure of phase space, spin-glass
order is destroyed in a field for short-range systems.

Finally, we also discuss extensions as well modifications of the
model to study different related problems in the physics of spin
glasses and present applications to algorithm development and testing.

\section{Model \& Numerical Method}
\label{sec:modelandobs}

We first introduce the one-dimensional Ising chain in detail and
explain its rich phase diagram. Furthermore, we describe exchange
(parallel tempering) Monte Carlo, a numerical method which is very
efficient to study spin-glass systems at low temperatures.

\subsection{The one-dimensional Ising chain}
\label{subsec:model}

\begin{figure}[t]
\includegraphics[width=0.50\textwidth]{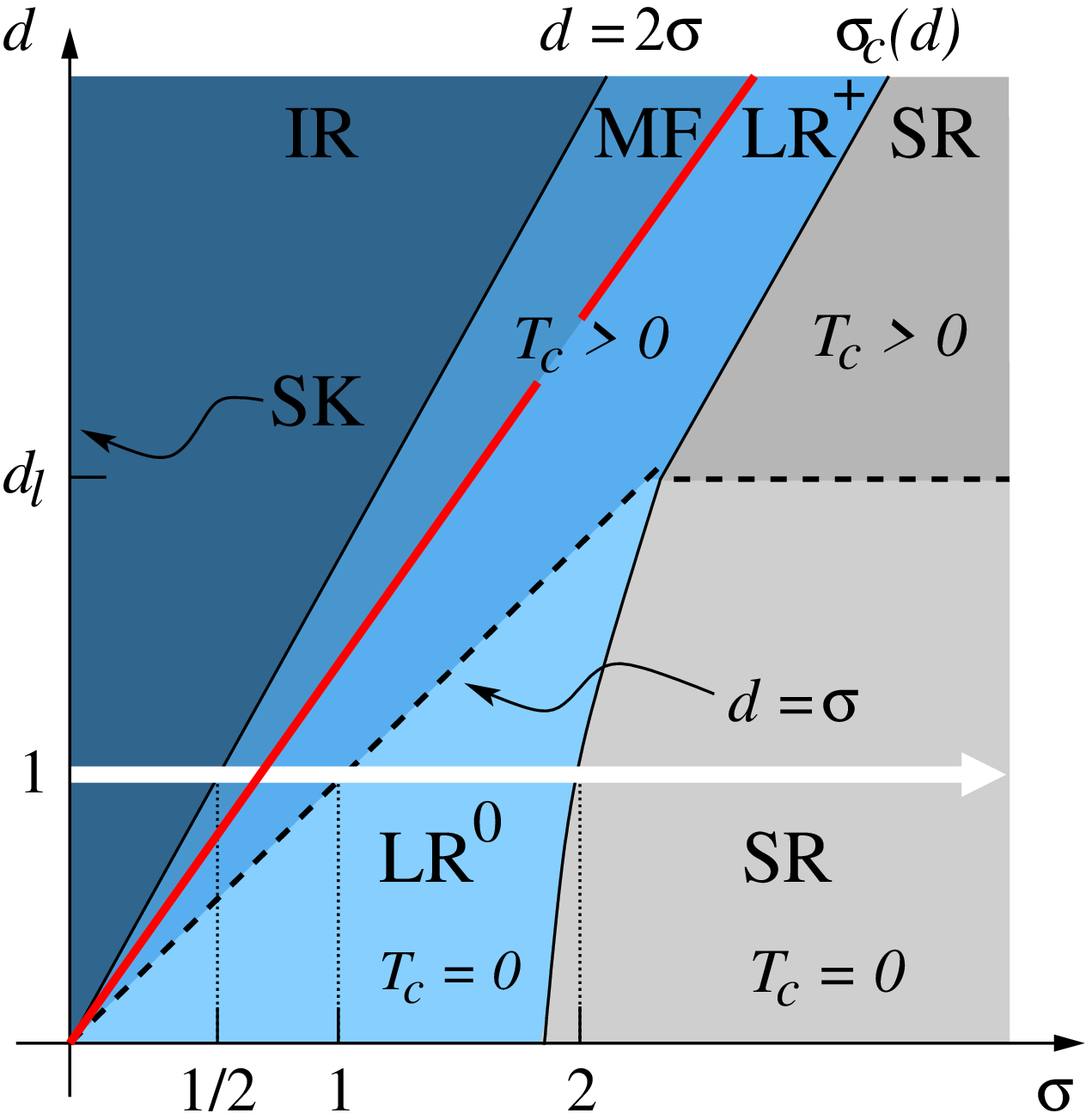}
\hspace*{0.20cm}
\includegraphics[width=0.46\textwidth]{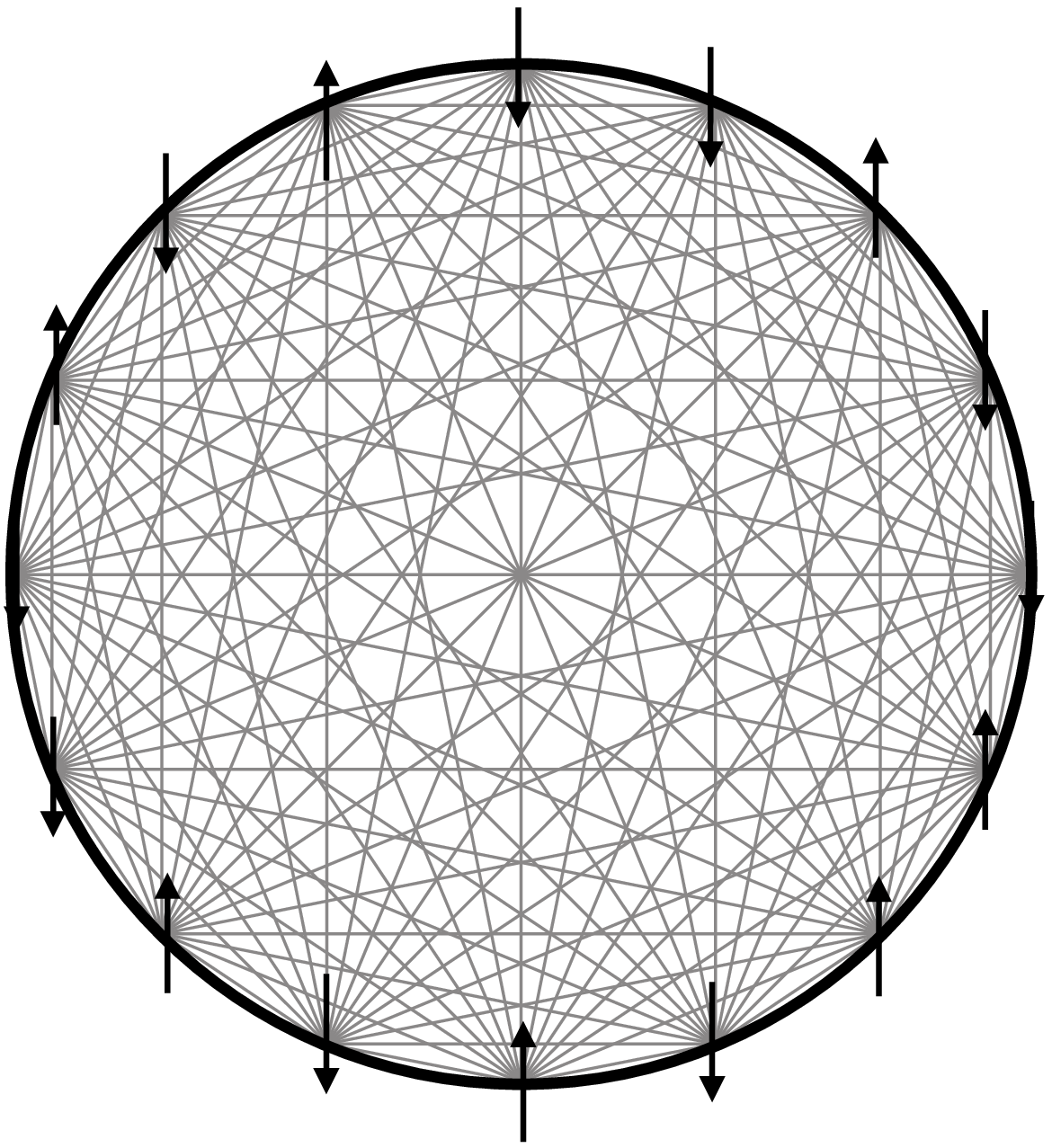}
\caption[]{
Left panel: Schematic phase diagram of the one-dimensional Ising
chain with power-law interactions \cite{fisher:88}. The white
horizontal arrow corresponds to $d = 1$. For $\sigma \le 1/2$ we expect
infinite-range (IR) behaviour reminiscent of the SK model. For $1/2 <
\sigma \le 2/3$ we have mean-field (MF) behaviour corresponding to
an effective space dimension $d_{\rm eff} \ge 6$, whereas for $2/3 <
\sigma \lesssim 1$ we have a long-range (LR$^+$) spin glass with a
finite ordering temperature $T_{\rm c}$. In these regimes $d_{\rm eff}
\approx 2/(2\sigma - 1)$ \cite{binder:86}.  For $1 \le \sigma \lesssim
2$ we have a long-range spin glass with $T_{\rm c} = 0$ (LR$^0$) and
for $\sigma \gtrsim 2$ the model displays short-range (SR) behaviour
with $T_c = 0$.  Figure adapted from Ref.~\cite{katzgraber:03}.
Right panel: Graphical representation of the one-dimensional Ising
chain with $L = 16$ spins.
}
\label{fig:dsigma}
\end{figure}

The Hamiltonian of the one-dimensional Ising chain with power-law
interactions \cite{kotliar:83,fisher:88,katzgraber:03} is given by
\begin{equation}
{\mathcal H}_{\rm 1D} = - \sum_{i<j} J_{ij} S_i S_j - \sum_i h_i S_i\, ,
\;\;\;\;\;\;\;
\;\;\;\;\;\;\;
\;\;\;\;\;\;\;
J_{ij}= c({\sigma}) \frac{\epsilon_{ij}}{{{r_{ij}}^\sigma}} \, ,
\label{eq:model}
\end{equation}
where $S_i \in\{\pm 1\}$ are Ising spins and the sum ranges over
all spins in the system. To ensure periodic boundary conditions,
the $L$ spins are placed on a ring, see Fig.~\ref{fig:dsigma}
(right panel). Here, $r_{ij} = (L/\pi)\sin(\pi |i - j|/L)$ is the
distance between the spins on the ring topology and $\epsilon_{ij}$ are
Gaussian-distributed random couplings of zero mean and standard
deviation unity. The constant $c(\sigma)$ is chosen such that
the mean-field transition temperature to a spin-glass phase is
$T_c^{\rm MF} = 1$; see Ref.~\cite{katzgraber:03} for details. The
model has a very rich phase diagram in the $d$--$\sigma$ plane, see
Fig.~\ref{fig:dsigma} (left panel). In this work we are interested in
$d = 1$ which corresponds to the thick horizontal white arrow in the
phase diagram. The universality class and range of the interactions
of the model can be continuously tuned by changing the power-law
exponent $\sigma$. Furthermore, there are theoretical predictions for
the critical exponents \cite{kotliar:83}: $\nu = 1/(2\sigma -1)$ for
$\sigma \le 2/3$, and $\eta = 3 - 2\sigma$.  Therefore, predictions
made for the mean-field spin-glass can be probed when the effective
space dimension (range of the interactions) is reduced. Furthermore,
because the efficiency of different algorithms often depends strongly
on the range of the interactions, the one-dimensional chain is an
ideal test bed to benchmark the efficiency of optimization algorithms.

In one space dimension, for $\sigma \le 1/2$ the model is in the
Sherrington-Kirkpatrick \cite{sherrington:75} infinite-range
universality class where the energy of the system needs to be
rescaled with the system size to avoid divergencies. In particular,
for $\sigma = 0$ the SK model is recovered exactly.  For $1/2 <
\sigma < 1$ the model has a finite-temperature spin-glass ordering
transition. Furthermore, for $1/2 < \sigma \le 2/3$ the system is in
the mean-field universality class corresponding to a high-dimensional
short-range spin-glass system above the upper critical dimension
$d_{\rm u} = 6$. For $2/3 < \sigma < 1$ the system is non-mean field,
whereas for $\sigma \ge 1$ the spin-glass phase only exists at $T =
0$, i.e., the lower critical dimension of short-range spin glasses
corresponds to $\sigma = 1$.

\subsection{Numerical method}
\label{subsec:numerical}

Because of a rough energy landscape and diverging relaxation times,
spin glasses are extremely difficult to study numerically. Any
numerical method used must have the potential to efficiently cross
energy barriers and thus sample the phase space evenly. Probably
one of the simplest, yet most efficient methods to study problems
with rough energy landscapes (beyond spin glasses) is the exchange
(parallel tempering) Monte Carlo method \cite{hukushima:96}.

The idea behind the method is to allow for a Markov process
in temperature space. $M$ copies of the system are simulated at
different temperatures, where the largest temperature is generally
chosen to be of the order of $2T_c^{\rm MF}$.  Besides the simple
Monte Carlo updates \cite{metropolis:49} on each spin of the
system, after a certain number of lattice sweeps the energies
of neighboring temperatures are compared and a Monte Carlo move
which swaps the temperatures of neighboring configurations is
proposed. With this approach, a configuration stuck in a metastable
state has the possibility to heat up and then cool back down to the
true equilibrium state thus effectively speeding up equilibration
by orders of magnitude. The position of the temperatures has to be
chosen with care: If neighboring temperatures are chosen too far
apart, a bottleneck in the temperature-space Markov process emerges
thus reducing the efficiency of the method. If the temperatures are
too close extra unnecessary overhead is introduced.  To select the
position of the temperatures, it is convenient to study the acceptance
probabilities of the global Monte Carlo moves.  Because in spin glasses
the susceptibility does not diverge, a generally good thumb-rule is to
select the position of the temperatures such that the probabilities are
between $0.2$ and $0.9$ and roughly independent of temperature. This is
not necessarily the case for other systems. We also refer the reader
to Ref.~\cite{katzgraber:06a} where an iterative feedback method is
introduced which ensures that the random walk of each configuration
in temperature space is optimal.

\begin{figure}[t]
\sidecaption
\includegraphics[width=0.55\textwidth]{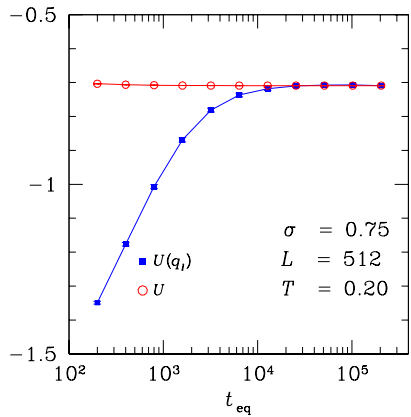}
\caption[]{
Equilibration test using Eq.~(\ref{eq:equil}). Once the energy $U$
computed directly and from the link overlap [$U(q_l)$] agree, the
system is in thermal equilibrium. Data for $L = 512$, $T = 0.20$
and $\sigma = 0.75$.
\vspace*{3.7cm}
}
\label{fig:equil}
\end{figure}

When using Gaussian-distributed disorder we test equilibration of
the Monte Carlo simulations by equating the link overlap $q_l$ to the
energy $U = -(1/N)\sum_{i,j}[J_{ij}\langle S_i S_j\rangle]_{\rm av}$
of the system \cite{katzgraber:01}, i.e.,
\begin{equation}
U(q_l) = \frac{(T_c^{\rm MF})^2}{2T}(q_l - 1)\;,
\;\;\;\;\;\;
\;\;\;\;\;\;
\;\;\;\;\;\;
q_l = \frac{2}{N}\sum_{i,j}\frac{[J_{ij}^2]_{\rm av}}{(T_c^{\rm MF})^2}
[\langle S_i S_j\rangle_T^2]_{\rm av} \, .
\label{eq:equil}
\end{equation}
In Eq.~(\ref{eq:equil}), $\langle \cdots \rangle_T$ represents
a thermal average and $[\cdots]_{\rm av}$ an average over the
disorder. $T$ is the temperature of the system. As can be seen
in Fig.~\ref{fig:equil}, starting from a random configuration will
underestimate $U(q_l)$, whereas the energy $U$ will be overestimated.
Once both agree, the system is in thermal equilibrium. Note that the
method can be easily extended to system with (Gaussian distributed)
external fields \cite{katzgraber:05c}.

\section{Selected results}
\label{sec:results}

We have applied the one-dimensional Ising chain to several problems
in the physics of spin glasses. Below we present in more detail two
questions which lie at the core of the applicability of the mean-field
solution to short-range spin glasses. In the following we compare
the mean-field SK model ($\sigma = 0$) to the one-dimensional Ising
chain for $\sigma = 0.75$ where the model is in the non-mean-field
universality class.

\subsection{Do spin glasses order in a magnetic field?}
\label{subsec:atline}

The applicability of spin-glass models to other fields of
science relies heavily on the existence of a spin-glass phase
in a field. Many mappings onto spin-glass models produce
external field terms. While the mean-field model has been
shown to have a spin-glass phase in a field, it has been unclear
until recently if short-range spin glasses order in a field as well
\cite{bhatt:85,ciria:93b,kawashima:96,billoire:03b,marinari:98d,houdayer:99,krzakala:01,takayama:04,young:04,jonsson:05a}.
Simulations of three-dimensional spin-glass models
\cite{young:04,joerg:08a} suggest that the de Almeida-Thouless line
\cite{almeida:78}, which separates the spin-glass from the paramagnetic
state in the $H$--$T$ phase diagram does not exist for realistic
short-range Ising spin glasses. Although the aforementioned studies
in three space dimensions using the finite-size two-point correlation
length \cite{ballesteros:00} provide clear evidence that short-range
spin glasses do not order in a field, they do not shed any light on
the behavior of short-range spin glasses with space dimensions above
the upper critical dimension.

In Ref.~\cite{katzgraber:05c} the one-dimensional Ising chain
has been studied in an externally applied Gaussian-distributed
random field---which has a similar behavior than a uniform field
---for different exponents $\sigma$ of the power-law interactions.
For exponents which correspond to effective space dimensions above
the upper critical dimension, a spin-glass state in a field is found,
whereas for exponents $\sigma > 2/3$ which correspond to effective
space dimensions less than six, no de Almeida-Thouless line could
be found for simulations down to very low temperatures.  Technical
details about the simulation, and in particular the parameters of
the simulation can be found in Ref.~\cite{katzgraber:05c}.

In order to probe the existence of a spin-glass state we add
an external (random) field to the Hamiltonian, i.e., ${\mathcal
H}_{\rm 1D} \rightarrow {\mathcal H}_{\rm 1D} - \sum_i h_i S_i$. The
reasons for using random fields are the ability to thoroughly
test for equilibration of the Monte Carlo method (for detail see
Refs.~\cite{young:04} and \cite{katzgraber:05c}). Furthermore,
exchange Monte Carlo performs better.

\begin{figure}[b]
\hspace*{-0.5cm}
\includegraphics[width=0.55\textwidth]{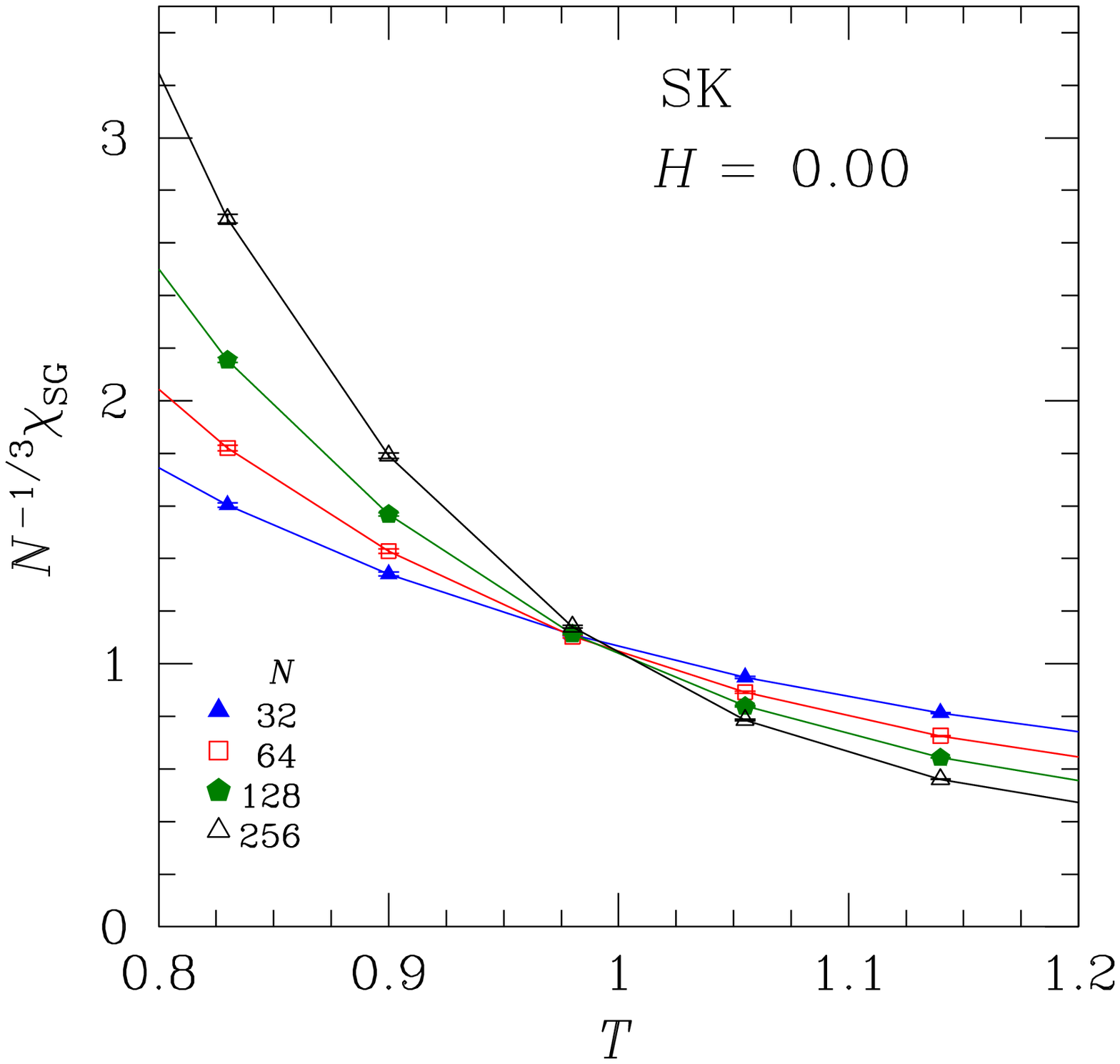}
\hspace*{-0.5cm}
\includegraphics[width=0.55\textwidth]{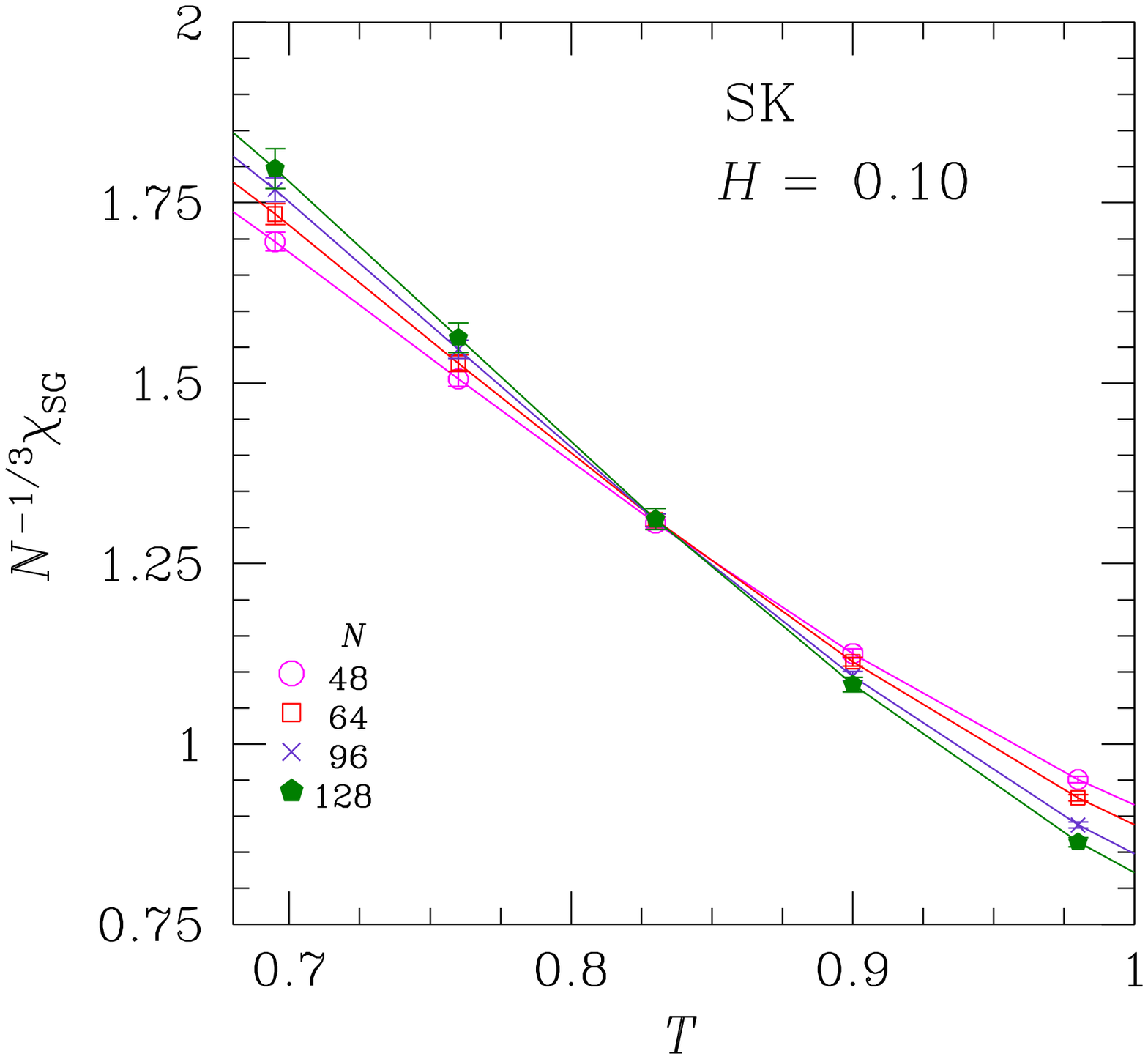}
\caption[]{
Left panel: Scaled spin-glass susceptibility $N^{-1/3}\chi_{\rm SG}$
as a function of temperature for the mean-field Sherrington-Kirkpatrick
model at zero external field. The data cross at $T_c(H=0) =  1.0$
in agreement with analytical results. Right panel: Same observable
and model as depicted in the left panel, except for $H = 0.10$. The
data cross at $T_c(H=0.10)\approx 0.82$ thus clearly showing that
the mean-field model orders in a field, as expected from theoretical
results.
}
\label{fig:at-0.00}
\end{figure}

To test for the existence of the transition for $\sigma > 1/2$, we
compute the finite-size correlation length from the Fourier transform
of the spin-glass susceptibility \cite{palassini:99b,ballesteros:00}:
\begin{equation}
\chi_{\rm SG}(k) = \frac{1}{N} \sum_{ij}\left[ \left( \langle S_i S_j
\rangle_T - \langle S_i \rangle_T\langle S_j \rangle_T\right)^2\right]_{\rm
av} e^{ik(R_i - R_j)} \, .
\label{eq:chisg}
\end{equation}
After performing an Ornstein-Zernicke approximation we obtain for
the two-point finite-size correlation length
\begin{equation}
\xi_L = \frac{1}{2\sin(k_{\rm min}/2)}
\left[ \frac{\chi_{\rm SG}(0)}{\chi_{\rm SG}(k_{\rm min})} - 1\right]^{1/(2\sigma -1 )}\, ,
\label{eq:xil}
\end{equation}
where $\chi_{\rm SG}(0)$ is the standard spin-glass susceptibility and
$k_{\rm min} = 2\pi/L$. The finite-size correlation length divided by
the system size is a dimensionless quantity which scales as $\xi_L/L
= \tilde{X}[L^{1/\nu}(T - T_c)]$.  Because in the infinite-range
universality class no correlation length can be computed, we exploit
the fact that the critical exponent $\eta = 1/3$ is exactly known
for the SK model \cite{billoire:07,joerg:08a}. Therefore, we locate
the transition in the SK model by studying $\chi_{\rm SG}/N^{1/3}
= \tilde{C}[L^{1/\nu}(T - T_c)]$, where $\chi_{\rm SG} = \chi_{\rm
SG}(k = 0)$.  Once the respective observables for different system
sizes cross we have a spin-glass state for $T \le T_c$, where $T_c$
is given by the crossing point.

\begin{figure}[t]
\hspace*{-0.5cm}
\includegraphics[width=0.55\textwidth]{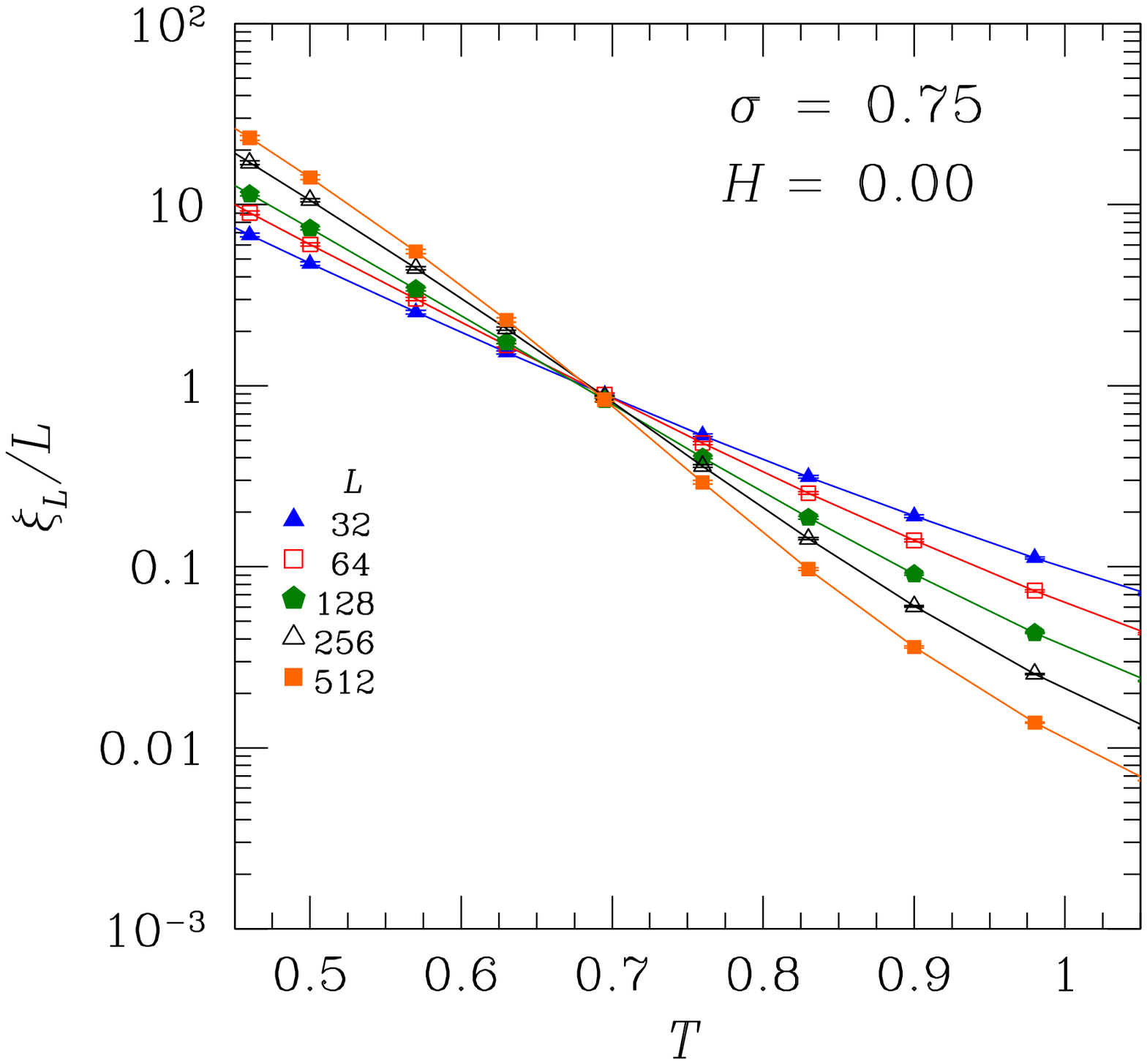}
\hspace*{-0.5cm}
\includegraphics[width=0.55\textwidth]{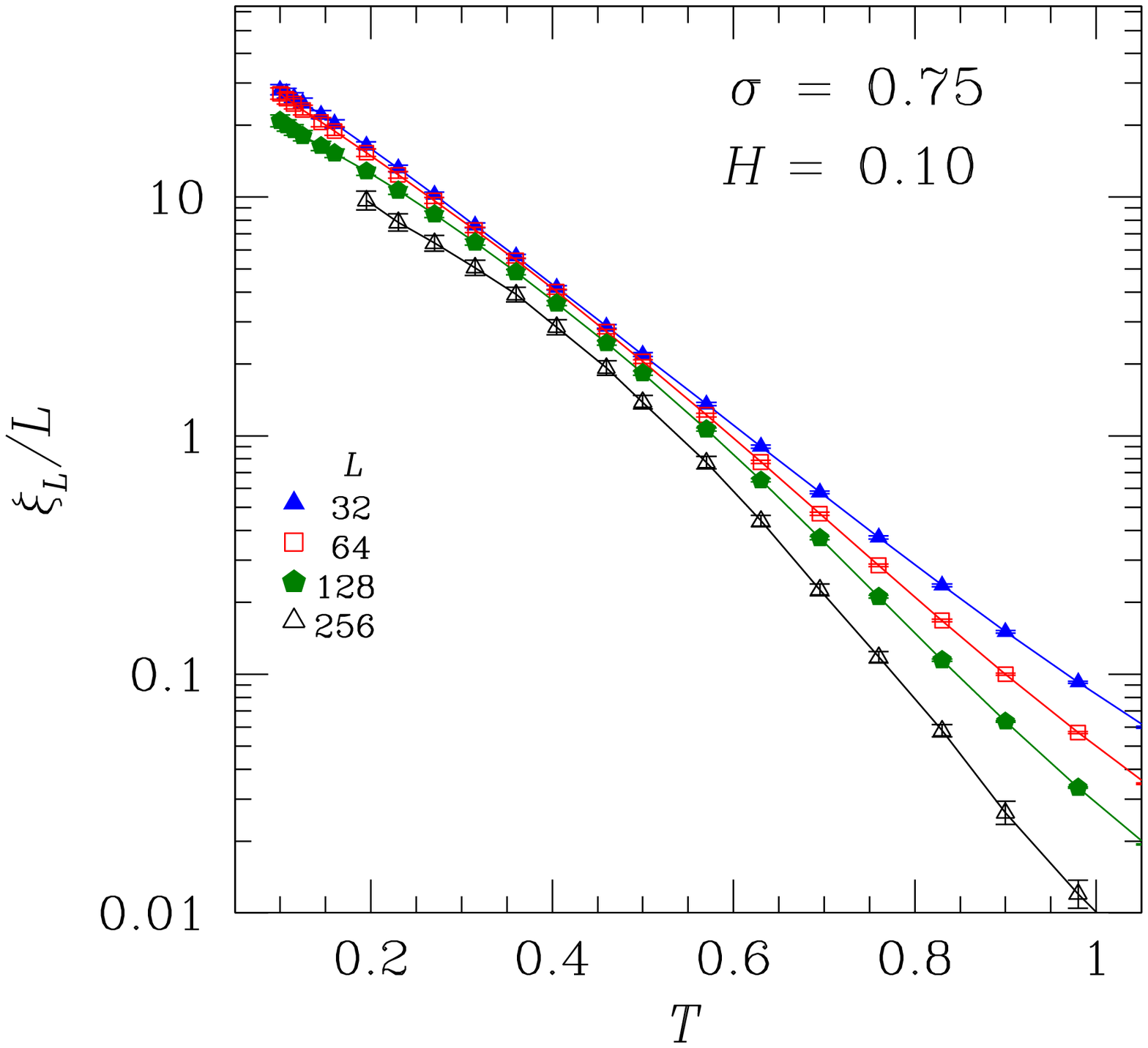}
\caption[]{
Left panel: Finite-size correlation length divided by the system
size as a function of temperature for the one-dimensional Ising
spin chain with $\sigma = 0.75$ at zero field. In this regime the
system is not in the mean-field universality class.  The data cross
cleanly at $T_c(H=0) \approx 0.69$.  Right panel: Same observable
and model as the left panel, except for $H = 0.10$. Note that for
temperatures as low as $T = 0.1$ there is no crossing visible,
suggesting that there is no spin-glass state in a field. Figure
adapted from Ref.~\cite{katzgraber:05c}.
}
\label{fig:at-0.75}
\end{figure}

In Fig.~\ref{fig:at-0.00} we show $\chi_{\rm SG}/N^{1/3}$ for the SK
model ($\sigma = 0$) as a function of temperature for zero, as well as
an external field of strength $H = 0.10$. In both cases the data cross,
indicative of a transition in zero as well as finite fields. This is
not the case for the one-dimensional model with $\sigma = 0.75$. While
the data of the finite-size correlation length at zero field clearly
show a transition at $T_c \approx 0.69$ (see Fig.~\ref{fig:at-0.75},
left panel), this is not the case for $H = 0.10$ where the data do
not cross even for temperatures considerably lower than the critical
temperature (see Fig.~\ref{fig:at-0.75}, right panel).

The presented results clearly show the numerical existence of an
AT line for the mean-field SK model, whereas for the model at
$\sigma = 0.75$ (outside the mean-field universality class there is
no sign of a transition in a small but finite field). Together with
results presented in Ref.~\cite{katzgraber:05c} we thus conclude that
short-range spin glasses below the upper critical dimension do not
order in an externally-applied magnetic field.

\subsection{Are spin glasses ultrametric?}
\label{subsec:ultra}

Ultrametricity is an intrinsic property of the Parisi solution of
the mean-field model \cite{mezard:84} and it can be described in the
following way: Consider an equilibrium ensemble of states at $T <
T_{\rm c}$ and pick three, $S^\alpha$, $S^\beta$ and $S^\gamma$,
at random. Order them so that their distances $d_{\alpha\beta}
=(1-q_{\alpha\beta})/2$, where $q_{\alpha\beta} = L^{-1}\sum S_i^\alpha
S_i^\beta$ is the spin overlap, satisfy $d_{\alpha\gamma}\geq
d_{\gamma\beta}\geq d_{\alpha\beta}$.  Ultrametricity means that in
the thermodynamic limit we obtain $d_{\gamma\beta}=d_{\alpha\beta}$
with probability $1$, i.e., the states lie on an isosceles triangle.

To date, the existence of ultrametricity for short-range
spin glasses---which would validate the applicability of
the mean-field solution to short-range systems---is highly
controversial. Recent results \cite{hed:03} suggest that short-range
systems are not ultrametric, whereas other opinions exist
\cite{franz:00,contucci:07,joerg:07}. Because the one-dimensional
Ising chain allows for tuning the system away from the mean-field
universality class, it presents itself as the ideal test-bed for this
problem. Below we present results for $\sigma = 0.0$ (SK) as well as
$0.75$ (non-mean-field regime) using an approach closely related to
the one used by Hed {\em et al.}~\cite{hed:03}.

\begin{figure}[t]
\includegraphics[width=0.325\textwidth]{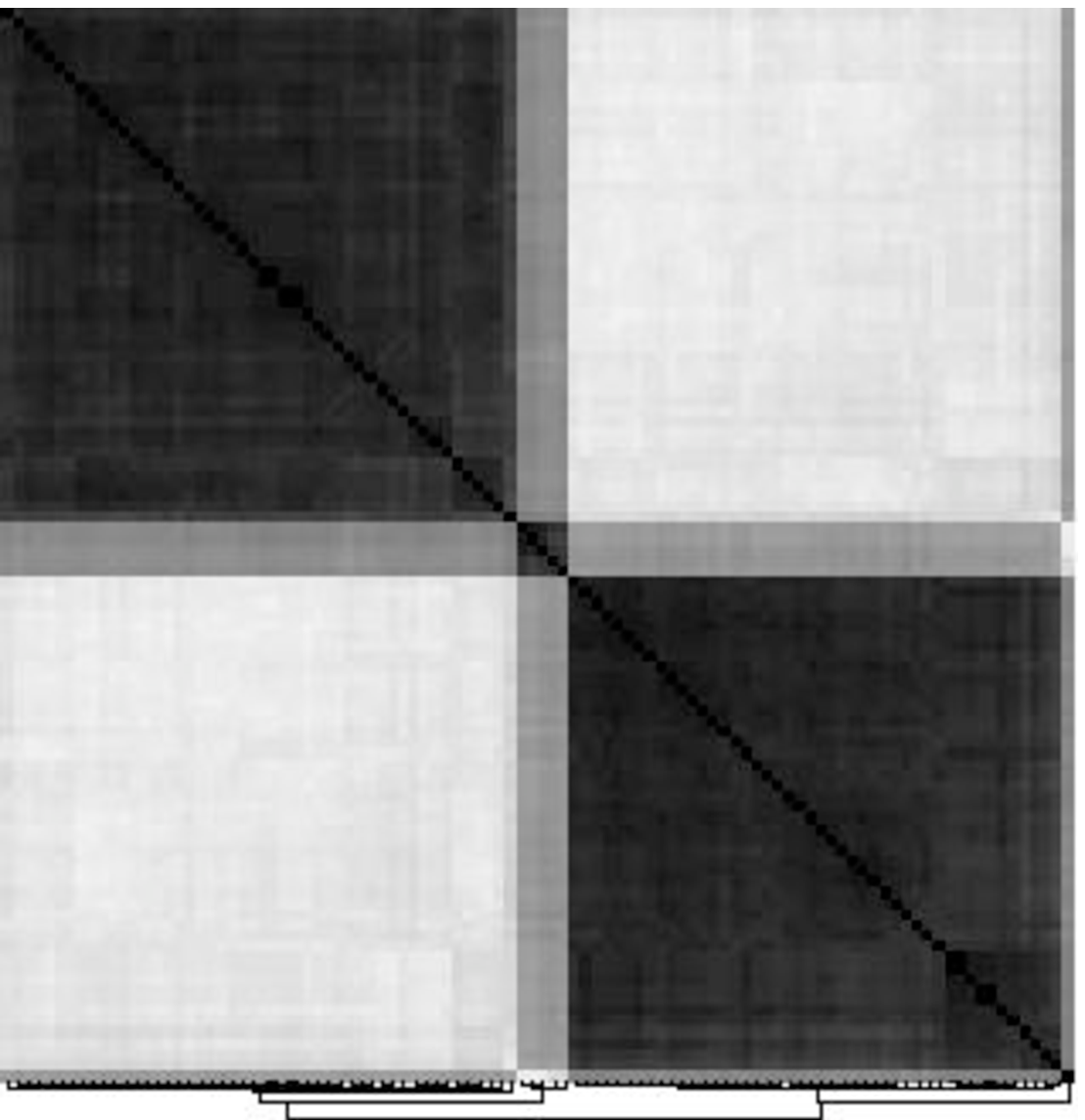}
\includegraphics[width=0.325\textwidth]{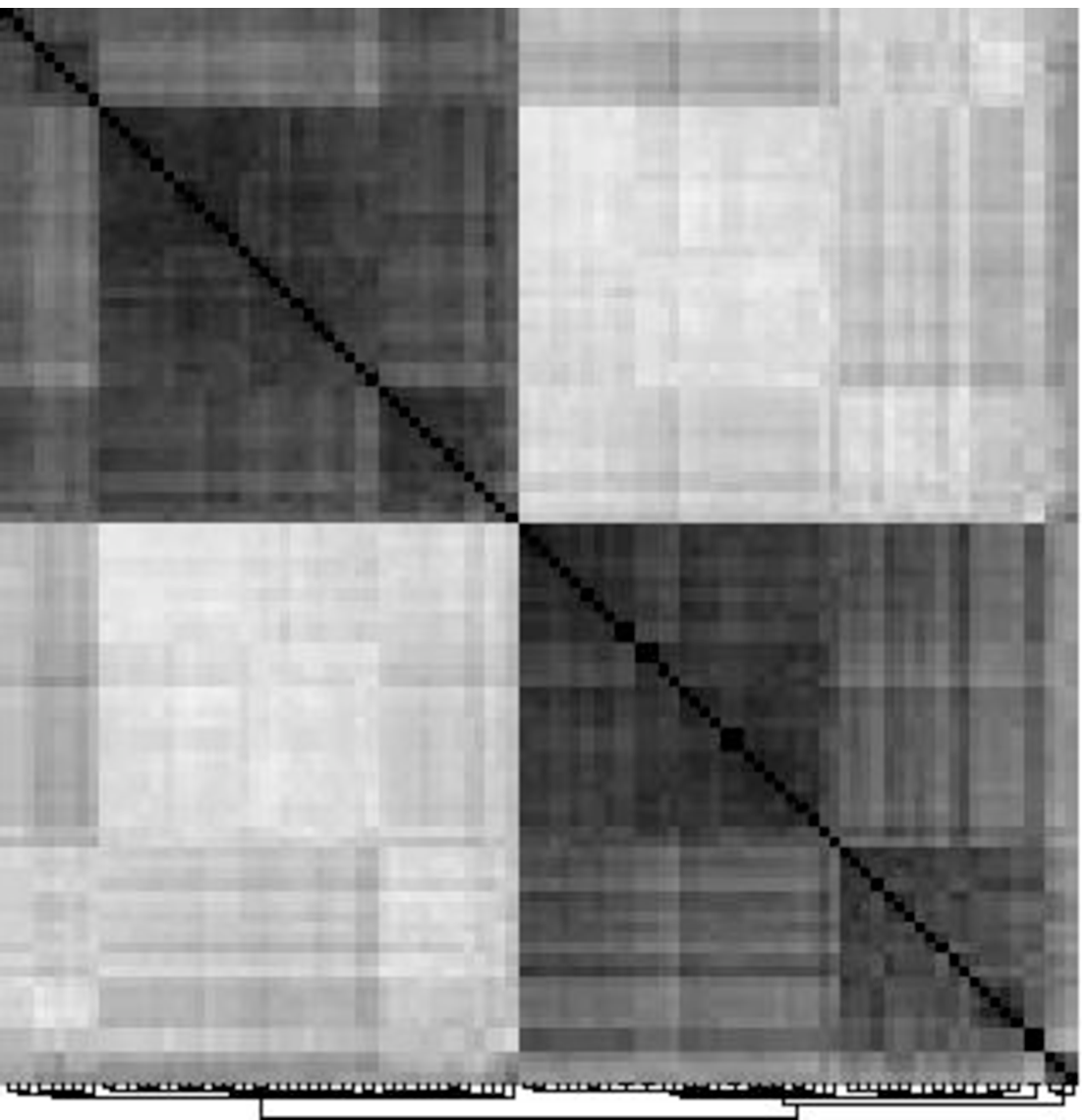}
\includegraphics[width=0.325\textwidth]{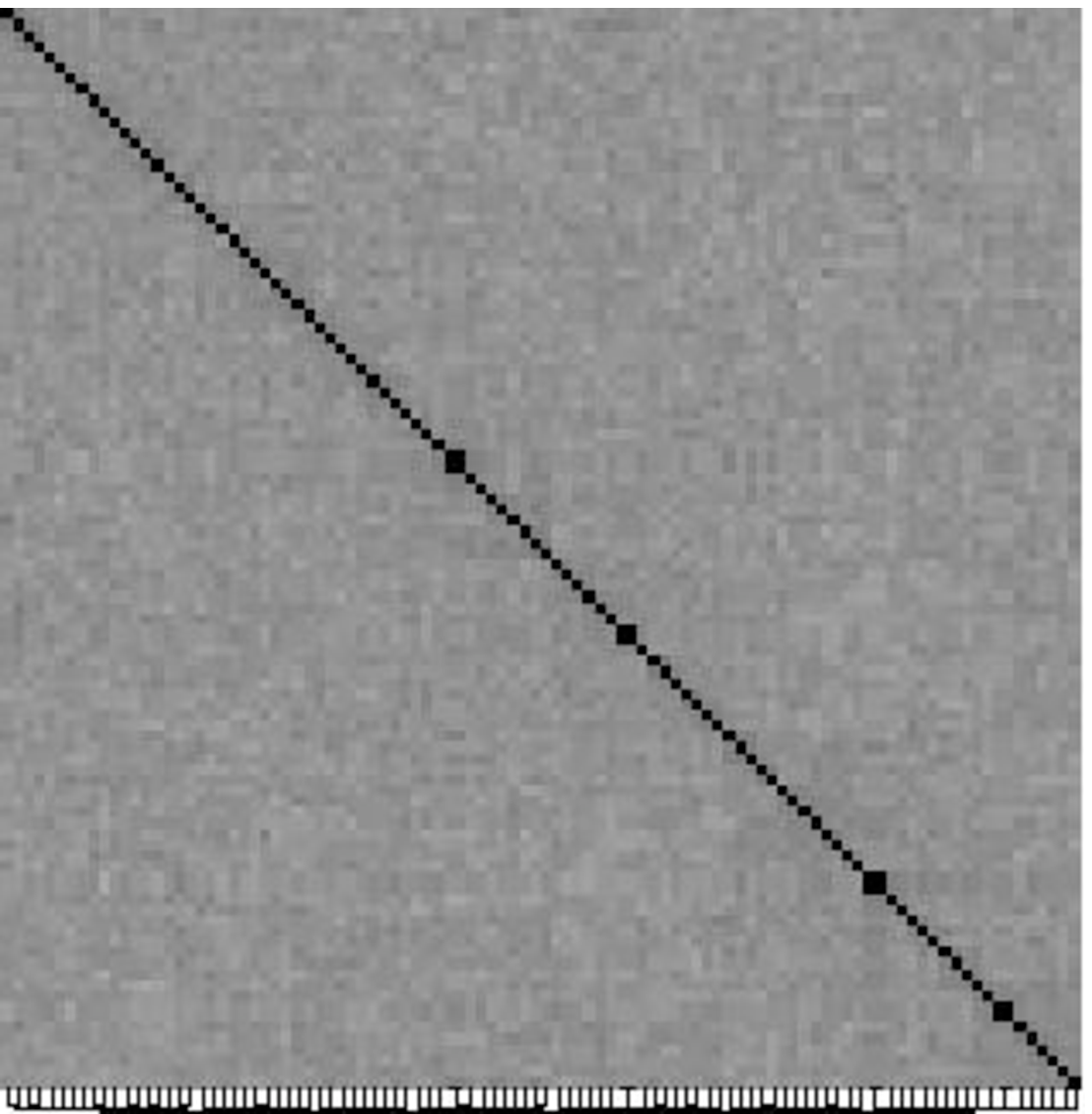}
\caption[]{
Dendrograms and distance matrices. Darker colors correspond to closer
distances in phase space. Left panel: SK model at $T = 0.4$ ($L =
1024$).  The distance matrix shows clear structure below $T_c$. Middle
panel: One-dimensional Ising chain for $\sigma = 0.75$ and $T = 0.40 <
T_c$ ($L = 512$). Again the data show structure. This is in contrast
to the right panel which shows data for the one-dimensional chain at
$T = 1.40 \gg T_c$ ($L = 512$, $\sigma = 0.75$).
}
\label{fig:dendros}
\end{figure}

We generate 1000 equilibrium states (spin configurations) for 1000 --
4000 disorder instances of the model using exchange Monte Carlo at
$T = 0.4T_c$ (i.e., $T = 0.4$ for the SK model and $T = 0.27$ for the
one-dimensional chain with $\sigma = 0.75$). The temperature used is
chosen such that we probe deep in the spin-glass phase, but not too low
to avoid trivial state triangles.  The generated states are in turn
sorted using Ward's hierarchical clustering approach \cite{ward:63}
(see Fig.~\ref{fig:dendros}).  The clustering procedure starts with $L$
clusters which contain one state and the two closest lying clusters
are merged.  Distances are measured in terms of the hamming distance
$d_{\alpha\beta} =(1-q_{\alpha\beta})/2$.  This procedure is repeated
until one large cluster is obtained. Once the states are clustered,
we select three states from different branches of the left sub-tree
(see Ref.~\cite{hed:03} for details) and sort the distances: $d_{\rm
max} \geq d_{\rm med} \geq d_{\rm min}$. We compute the correlator
\begin{equation} 
K = \frac{d_{\rm max} - d_{\rm med}}{\varrho(d)} ,
\label{eq:K}
\end{equation}
where $\varrho(d)$ is the width of the distribution of distances.
If the space is ultrametric, we expect $d_{\rm max} = d_{\rm med}$
for $L \rightarrow \infty$. This means for the distribution $P(K)
\rightarrow \delta(K = 0)$ for $L \rightarrow \infty$.

\begin{figure}[t]
\hspace*{-0.5cm}
\includegraphics[width=0.55\textwidth]{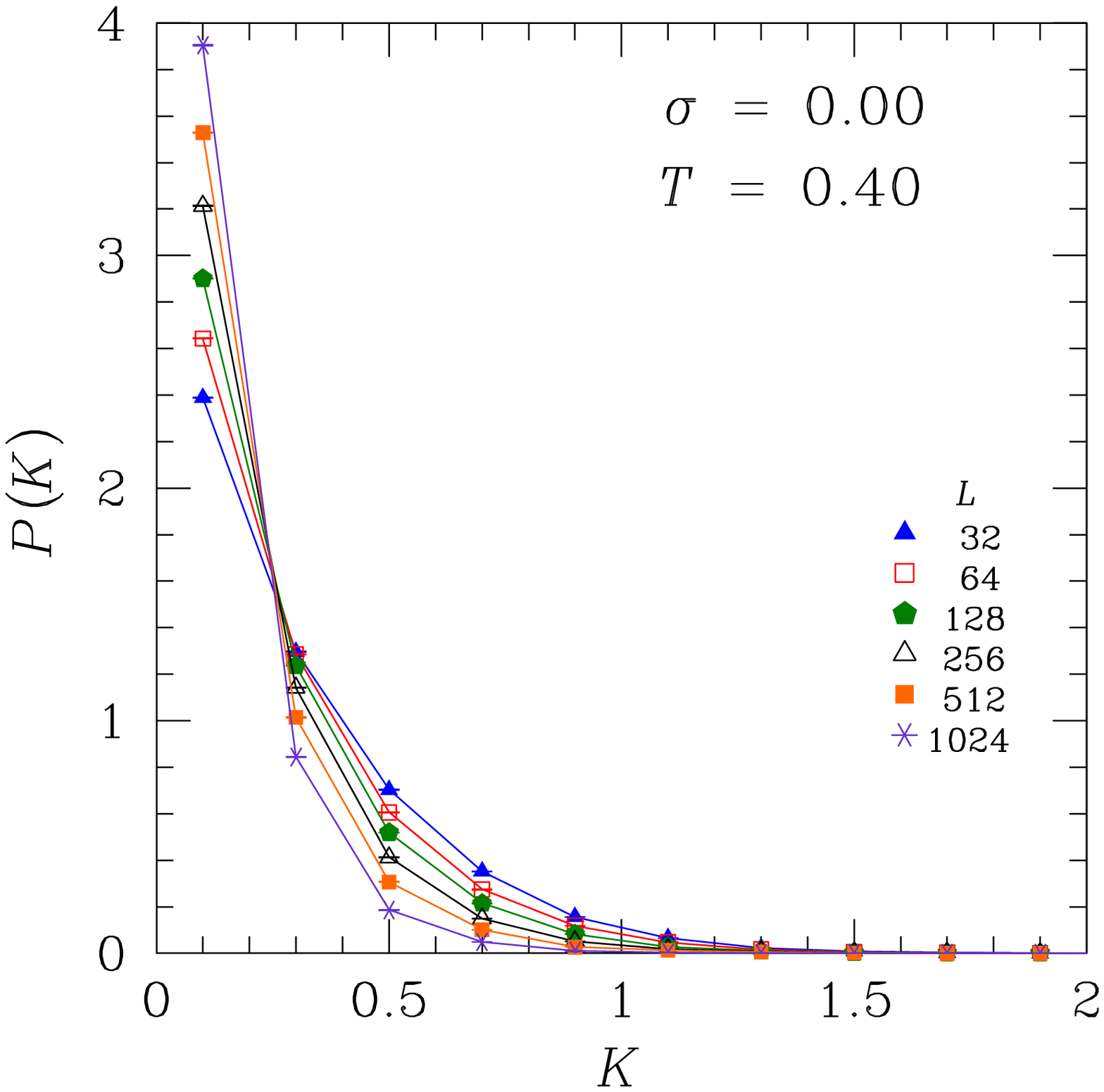}
\hspace*{-0.5cm}
\includegraphics[width=0.55\textwidth]{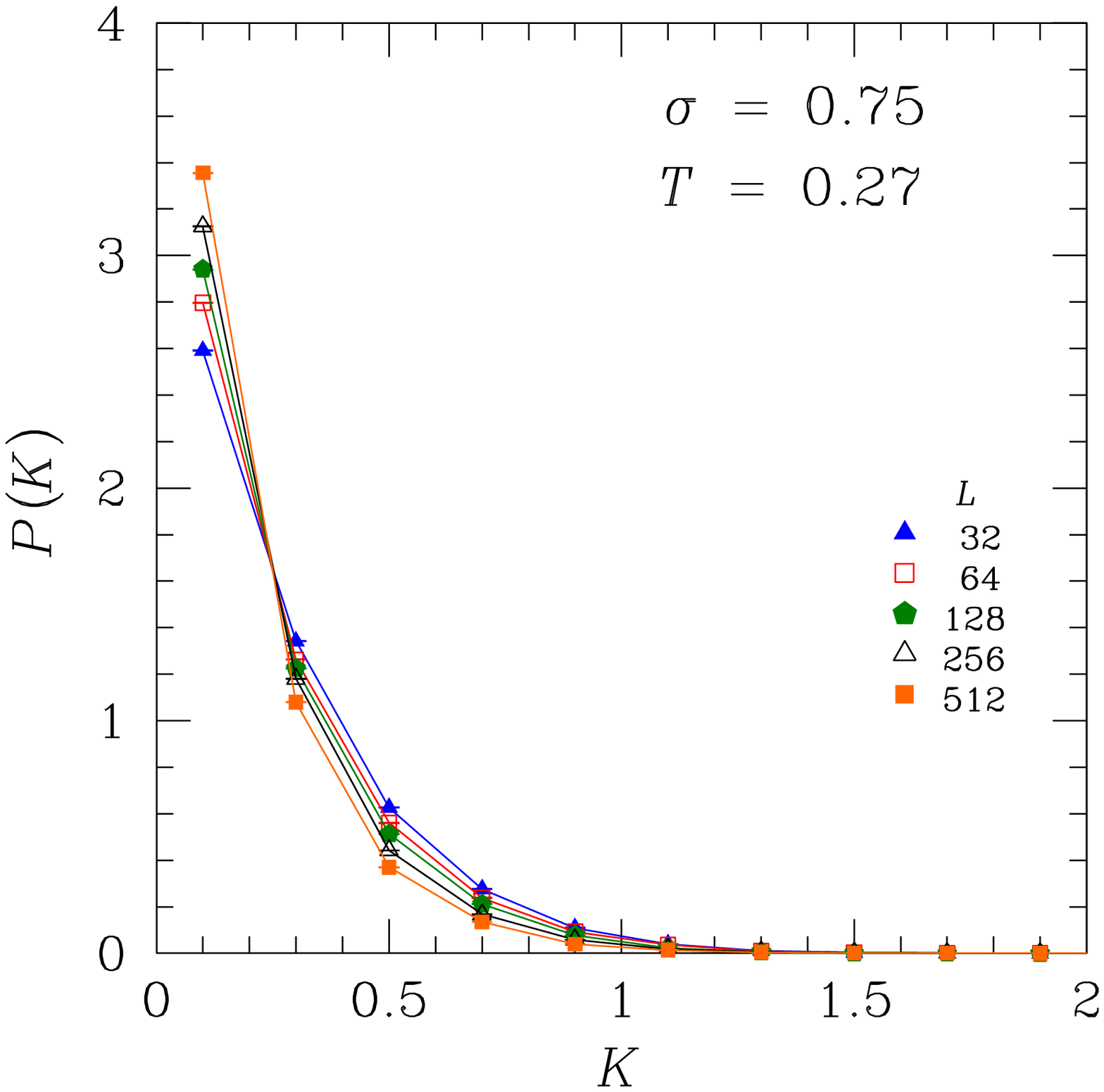}
\caption[]{
Left panel: Distribution $P(K)$  for the mean-field SK model at $T =
0.4T_c$. The data peak for $K \rightarrow 0$ with increasing system
size showing clearly that phase space is ultrametric.  Right panel:
Same observable as in the left panel for the one-dimensional Ising
chain with $\sigma = 0.75$ (non-mean-field universality class) at $T
= 0.27 \approx 0.4T_c$. While the divergence at $K = 0$ is less
pronounced, the data show a similar behavior than in the left panel.
}
\label{fig:ultra}
\end{figure}

In Fig.~\ref{fig:ultra} (left panel) we show data for the distribution
$P(K)$ for the SK model at $T = 0.4T_c$. For increasing system size
the data seem to converge to a limiting delta function. This is not the
case for $T = T_c$ (not shown) where the data are independent of system
size and show no divergence for $K \rightarrow 0$.  This suggests that
the used observable correctly captures the underlying ultrametric
behavior.  Furthermore, studies of cophonetic distances show that
the structures found in the dendrograms are not arbitrary.  Figure
\ref{fig:ultra} (right panel) shows $P(K)$ for the one-dimensional
Ising chain with $\sigma = 0.75$ at $T = 0.27 \approx 0.4T_c$ for a
range of system sizes. The data show a similar behavior than for the SK
model, although the effect is not as pronounced. Further simulations
at $\sigma$ values larger than $0.75$ as well as a quantitative study
of the number of clusters and RSB layers shall clarify with certainty
if short-range spin glasses have an ultrametric phase structure or not.

\section{Future directions}
\label{sec:future}

In the past, we have studied several properties of spin glasses using
the one-dimensional Ising chain, such as the nature of the spin-glass
state \cite{katzgraber:03,katzgraber:03f,katzgraber:05d}, ground-state
energy distributions of spin glasses \cite{katzgraber:04c}, the
existence of a spin-glass state in a field \cite{katzgraber:05c}
(see above), field chaos in spin glasses \cite{katzgraber:08},
local-field distributions in spin glasses \cite{boettcher:07},
as well as ultrametricity in spin glasses \cite{katzgraber:08}
(see above). Furthermore, other groups have also studied other open
questions in the physics of spin glasses with this model, such as
nonequilibrium problems \cite{montemurro:03} or different cumulants
of the order parameter distribution \cite{leuzzi:99}.  All previous
studies had been done on the model presented in Eq.~(\ref{eq:model})
using Ising spins. In this section we mention some extensions,
as well as modifications of the model which can be used to study
different problems.

\subsection{Variations on the model}
\label{subsec:variations}

Recently, a one-dimensional spin-glass chain with
Heisenberg spins has been studied in Ref.~\cite{matsuda:07}
to test the controversial spin-chirality decoupling scenario
\cite{kawamura:92,lee:03,hukushima:05,campos:06} proposed by Kawamura.
It is unclear to date what the nature of the spin-glass state in
Heisenberg spin glasses is. In particular, it is unclear if spin
and chirality degrees of freedom decouple. To test this scenario,
simulations of the one-dimensional Heisenberg chain \cite{matsuda:07}
at $\sigma = 1.1$ have been performed. For $\sigma = 1.1$ the spin
degrees of freedom only order at $T = 0$, whereas results suggest that
the chirality degrees of freedom order at finite nonzero temperatures.
Similar studies could be performed for models with planar XY spin
degrees of freedom, as well as Potts spins (work in progress).

Finally, the Hamiltonian can also be modified to include, for example,
$p$-spin interactions to study structural glasses \cite{moore:06}
(work in progress). Preliminary results suggest that the model has
a finite ordering temperature in the mean-field regime. 

\subsection{Studying larger systems with dilution}
\label{subsec:dilution}

While the linear system sizes $L$ studied with the one-dimensional
Ising chain are considerably larger than the system sizes accessible in
short-range systems, the fact that the model is fully-connected makes
it difficult to access large numbers of spins because any algorithm
would have to do ${\mathcal O}(L^2)$ updates at every Monte Carlo
sweep. This is because the system has ${\mathcal O}(L^2)$ interactions
between the spins. Recently, Leuzzi and collaborators suggested a
variation of the model which is diluted, thus drastically reducing the
number of neighbors for each spin \cite{leuzzi:08}. In their version,
a random bimodally-distributed bond between two spins is placed
with a power-law dependent probability adjusted such that the mean
connectivity $z$ is always $6$ for all $\sigma$. This has the effect
that for $\sigma \rightarrow 0$ we recover the Viana-Bray model with
fixed connectivity \cite{viana:85}. Because of the dilution, systems
of $10^4$ spins can be studied to temperatures as low as $\sim 0.4T_c$.

\begin{figure}[b]
\hspace*{-0.5cm}
\includegraphics[width=0.55\textwidth]{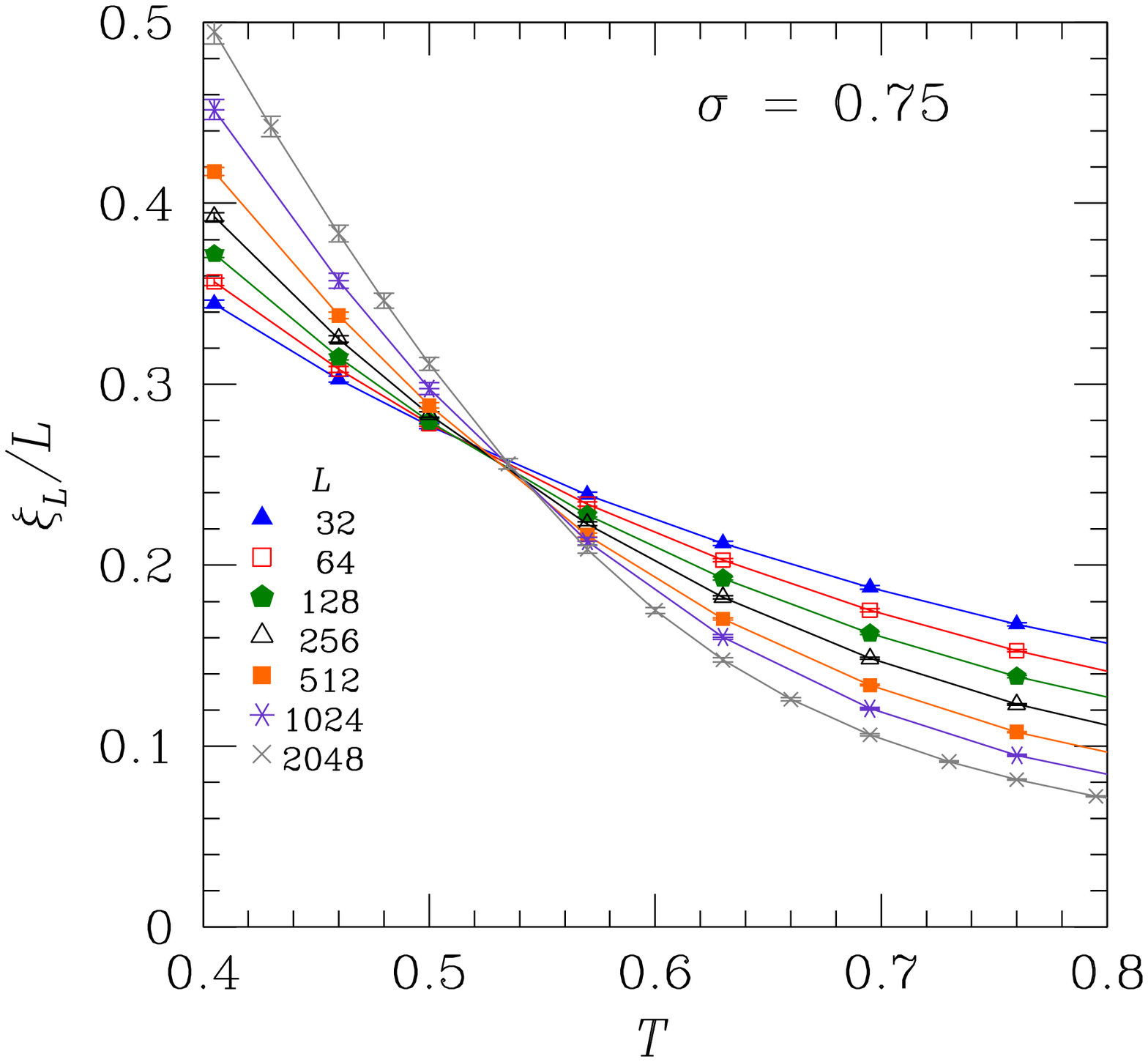}
\hspace*{-0.5cm}
\includegraphics[width=0.55\textwidth]{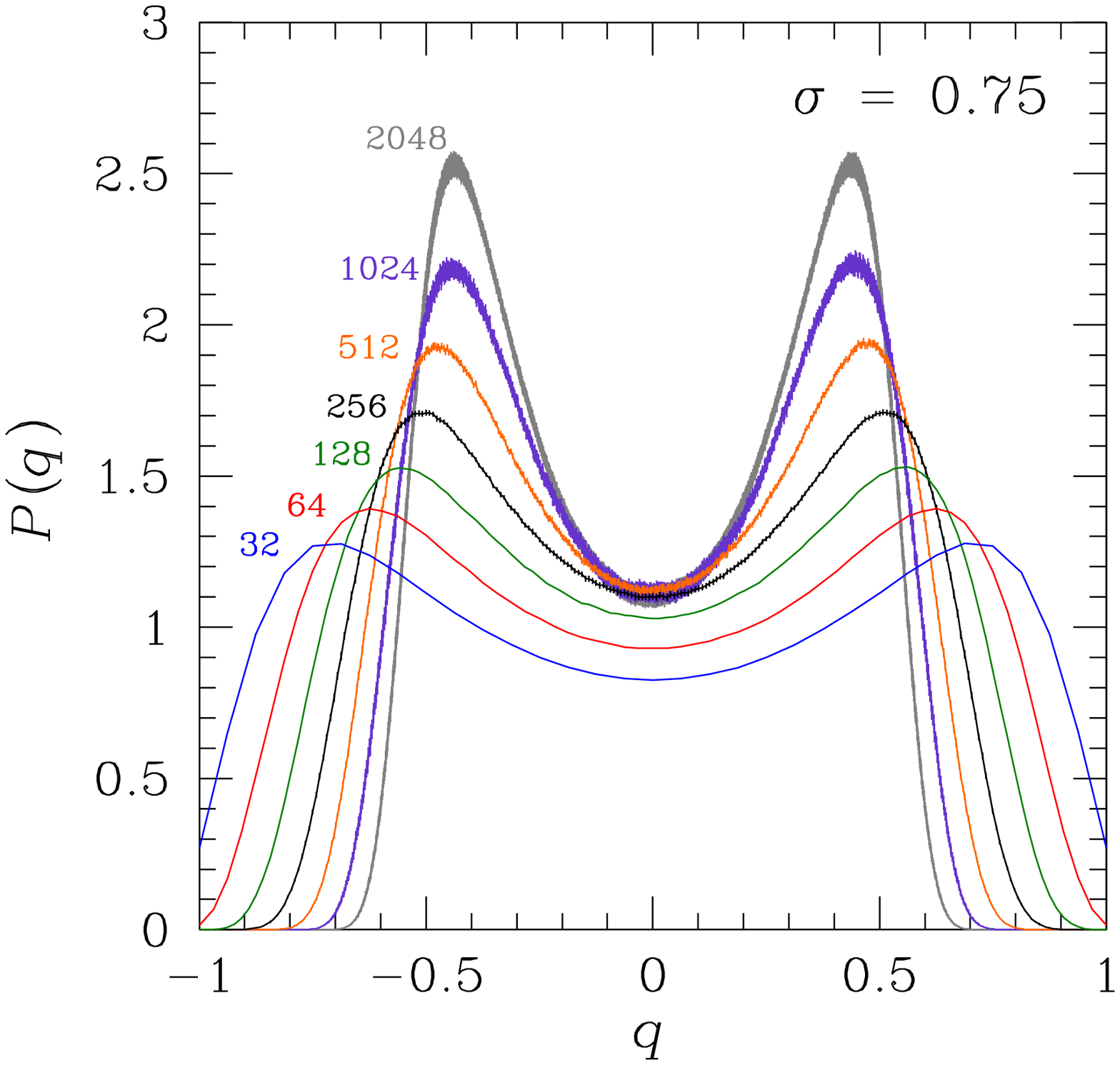}
\caption[]{
Left panel: Finite-size correlation length for the power-law diluted
one-dimensional Ising spin glass with variable connectivity. The data
cross at $T_c \approx 0.54$ illustrating the existence of a transition.
Right panel: Distribution of the spin overlap function $P(q)$ at $T =
0.4$ for different system sizes.  The width of the lines corresponds
to the error bars.
}
\label{fig:parisi}
\end{figure}

In Fig.~\ref{fig:parisi} we present data for a diluted system with
Gaussian-distributed random interactions and $\sigma = 0.75$. In this
case, the probability to place a bond between two spins is ${\mathcal
P}(J_{ij}\neq 0) = r^{-2\sigma}$, where $r$ is the distance between
the spins. The mean connectivity $z$ of the model is then given by $z
= 2\zeta(2\sigma)$ in the thermodynamic limit, where $\zeta$ is the
Riemann zeta function.  For $\sigma = 0$ we recover the SK model,
whereas, for example, for $\sigma = 0.75$ the mean connectivity is
only $z \approx 5.22$ thus allowing the study of large systems (note
that the interactions are rescaled such that $T_c^{\rm MF} = 1$).
In the left panel of Fig.~\ref{fig:parisi} the finite-size correlation
length as a function of temperature is shown. The data cross at $T_c
\approx 0.54$ signaling the existence of a spin-glass transition.
In the right panel of  Fig.~\ref{fig:parisi} we show the distribution
of the spin overlap $q = L^{-1}\sum_i S_i^\alpha S_i^\beta$. While
the data show corrections due to critical fluctuations, they converge
to a seemingly system-size independent value around $|q| \approx
0$. This would agree with the replica symmetry breaking scenario
by Parisi \cite{parisi:79,parisi:80,parisi:83,mezard:87} although
lower temperatures are needed to properly address this question.
Current work focuses on revisiting the existence of a spin-glass
state in a field using the model with dilution.

\section{Benchmarking of algorithms}
\label{sec:benchmarking}

\begin{figure}[b]
\sidecaption
\includegraphics[width=0.55\textwidth]{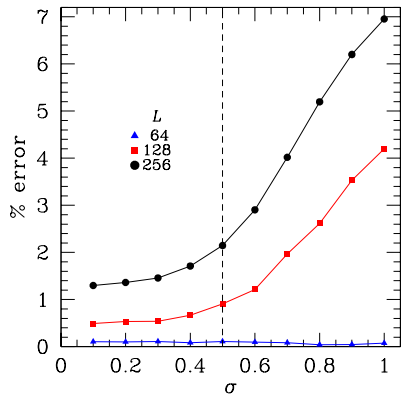}
\caption[]{
Percentage error in the ground-state energies obtained with hysteretic
optimization in comparison to exact ground states as a function
of the exponent $\sigma$ for different system sizes $L$.  Clearly,
the algorithm works best for $\sigma \lesssim 1/2$ (vertical dashed
line), i.e., in the infinite-range regime. Outside the infinite-range
regime, avalanches do not percolate the system and thus the algorithm
is less efficient.  Figure adapted from reference \cite{goncalves:08}.
\vspace*{0.6cm}
}
\label{fig:heo}
\end{figure}

Benchmarking optimization algorithms \cite{hartmann:01,hartmann:04}
plays a crucial role in the field of disordered and complex systems, as
well as many other interdisciplinary applications.  Knowing the range
of applicability of a given algorithms can be of great importance when
trying to solve a given problem.  For example, whereas the branch, cut
\& price algorithm \cite{barahona:88,liers:04,juenger:95} works best
for short-range systems, it is least efficient when the interactions
are long range \cite{katzgraber:04c}.

Recently, the hysteretic optimization heuristic \cite{pal:06} has
been introduced to estimate ground states of spin-glass systems. The
method is known to work well for the mean-field SK model, as well
as the traveling salesman problem \cite{hartmann:04}. The idea
behind hysteretic optimization is successive demagnetization at zero
temperature.  With additional shake-ups (field increases to further
randomize the system) close-to-ground-state configurations can be
obtained.  Recently, Gon{\c c}alves and Bottcher \cite{goncalves:08}
have studied the efficiency of the method on the one-dimensional
Ising chain. Data adapted from their work shown in Fig.~\ref{fig:heo}
clearly show that the method works best for infinite-range models
($\sigma \le 1/2$) where avalanches in the hysteresis loops proliferate
easily. While the error in finding the ground states increases slightly
with system size for $\sigma \lesssim 1/2$ the increase is considerably
stronger for larger values of $\sigma$.  As soon as the system is not
infinite-ranged, avalanches are small and the method is not efficient.

\section{Concluding remarks}
\label{sec:conclusions}

By using a one-dimensional spin-glass model with random power-law
interactions we have been able to shed some light on some of the
open questions in the physics of spin glasses.  The one-dimensional
spin-glass chain has the advantage over conventional models that
large linear system sizes can be studied. Furthermore, by changing
the power-law exponent of the interactions, different universality
classes ranging from the mean-field to the short-range universality
class can be probed. The latter feature of the model allows also for
efficient benchmarking of optimization algorithms.

\section*{Acknowledgments}

We would like to thank Stefan B\"ottcher, Ian Campbell, Thomas J\"org,
Florent Krz\c{a}ka{\l}a, Wolfgang Radenbach, David Sherrington and
Gergely Zimanyi for discussions.  In particular, we would like to
thank B.~Gon{\c c}alves and S.~B\"ottcher for sharing their data
from Ref.~\cite{goncalves:08}.  The simulations have been performed
on the Asgard, Brutus, Gonzales and Hreidar clusters at ETH Z\"urich.
H.G.K~acknowledges support from the Swiss National Science Foundation
under Grant No.~PP002-114713. A.P.Y.~acknowledges support from the
National Science Foundation under Grant No.~DMR 0337049.

\bibliography{refs}
\addcontentsline{toc}{section}{References}

\end{document}